\documentclass[artice,aps,pra,showpacs,twocolumn,superscriptaddress]{revtex4}
\usepackage{graphicx,color}
\usepackage{amsmath,amsthm,amsfonts,amssymb,bm}
\usepackage{times}
\usepackage{epsf}
\usepackage[colorlinks={true}]{hyperref}
\usepackage[T1]{fontenc}
\usepackage[utf8]{inputenc}

\hypersetup{citecolor={blue}, filecolor={blue}, linkcolor={blue}, urlcolor={blue}}

\newcommand{\commentold}[1]{}
\DeclareMathSymbol{:}{\mathpunct}{operators}{"3A}

\begin{document}
\title{Tightening the entropic uncertainty bound in the presence of quantum memory}
\author{F. Adabi}
\affiliation{Department of Physics, University of Kurdistan, P.O.Box 66177-15175 , Sanandaj, Iran}
\author{S. Salimi}
\email{shsalimi@uok.ac.ir}
\affiliation{Department of Physics, University of Kurdistan, P.O.Box 66177-15175 , Sanandaj, Iran}
\author{S. Haseli}
\affiliation{Department of Engineering Physics, Kermanshah University of Technology, Kermanshah, Iran}
\date{\today}
\begin{abstract}
The uncertainty principle is a fundamental principle in quantum physics. It implies that the measurement outcomes of two incompatible observables can not be predicted simultaneously. In quantum information theory, this principle can be expressed in terms of entropic measures. Berta \emph{et al}. [\href{http://www.nature.com/doifinder/10.1038/nphys1734}{ Nature Phys. 6, 659 (2010) }] have indicated that uncertainty bound can be altered by considering a particle as a quantum memory correlating with the primary particle. In this article, we obtain  a lower bound for entropic uncertainty in the presence of a quantum memory by adding an additional term depending on Holevo quantity and mutual information. We conclude that our lower bound will be tighten with respect to that of Berta \emph{et al.}, when the accessible information about measurements outcomes is less than the mutual information of the joint state. Some examples have been investigated for which  our lower bound is tighter than the Berta's \emph{et al.} lower bound. Using our lower bound, a lower bound for the entanglement of formation  of bipartite quantum states has obtained, as well as an upper bound for the regularized distillable  common randomness.
\end{abstract}
\maketitle

\section{Introduction}
The uncertainty principle is the most basic feature of quantum mechanics, which can be called the heart of quantum mechanics \cite{Heisenberg,Wheeler}. This principle bounds the uncertainties of measurement outcomes of two incompatible observables on a system in terms of the expectation value of their commutator. According to this principle, if measurement on a particle  is selected from a set of two observables $\lbrace X , Z  \rbrace $; then,  we have the following relation for quantum state $ \vert \psi \rangle$ \cite{Robertson}
\begin{equation}
\Delta X \Delta Z \geq \frac{1}{2}\vert \langle \psi \vert \left[ X , Z \right] \vert \psi \rangle \vert,
\end{equation}
where $\Delta X=\sqrt{\langle \psi \vert X^{2} \vert \psi \rangle - \langle \psi \vert X \vert \psi \rangle ^{2}} $ , $\Delta Y=\sqrt{\langle \psi \vert Y^{2} \vert \psi \rangle - \langle \psi \vert Y \vert \psi \rangle ^{2}} $ are the standard deviations and $\left[ X , Z \right]  = XZ -ZX $ is the commutator of the observables $X$ and $Z$. The uncertainty principle can be characterized in terms of Shannon entropies of the measurement outcomes probability distributions of the two observables. The most famous version of entropic uncertainty relation (EUR) was conjectured by Deutsch \cite{Deutsch}. It was improved by Kraus \cite{Kraus} and then proved by Maassen and Uffink \cite{Maassen}. It states that, given two observables $X$ and $Z$ with eigenbases $\{\vert x_{i}\rangle\}$ and $\{\vert z_{j}\rangle\}$, for any state $\rho_{A}$,
\begin{equation}\label{avali}
H(X)+H(Z)\geq  \log_{2}\frac{1}{c}=:q_{MU},
\end{equation}
where $q_{MU}$ is incompatibility measure,  $H(O)=-\Sigma_{k} p_{k}\log_{2}p_{k}$ is the Shannon entropy of the measured observable $O \in \lbrace X , Z \rbrace $, $p_{k}$ is the probability of the outcome $k$,  $c=\max\limits_{i,j}c_{ij}$,  and \hspace{5pt}$c_{ij}=\vert \langle x_{i} \vert z_{j} \rangle \vert^{2}. $

 Various attempts have been made to improve and to generalize this relation \cite{Bialynicki,Berta,Pati,Ballester,Vi,Wu,Wehner,Rudnicki,Rudnicki1,Pramanik,Maccone,Pramanik1,Zozor,Coles,Liu,Kamil,Zhang,R}. In the following, we explain  the generalization of EUR to the case in the presence of memory particle \cite{Berta}. One can describe the uncertainty principle by means of an interesting game between Alice and Bob. First, Bob prepares a particle in a quantum state and sends it to Alice. Then, Alice and Bob reach an agreement about measuring of two observables $X$ and $Z$ by Alice on the particle.  Alice does her measurement on the quantum state of the particle with one of the measurements and declares her choice of measurement to Bob. If Bob guesses the measurement outcome  correctly, he will win the game. The minimum of Bob's uncertainty about  Alice's measurement outcomes is bounded by Eq.(2). So far, it was assumed that there is just one particle, but if Bob prepares a correlated bipartite state $\rho_{AB}$ and sends just one of the particles to Alice and keeps the other particle as a quantum memory by himself, he can guess the Alice's measurement outcomes with a better accuracy. The uncertainty principle in the presence of  quantum memory has been studied by Berta \emph{et al}. \cite{Berta}, and they obtained the following relation
\begin{equation}\label{berta}
S(X \vert B)+S(Z \vert B) \geq q_{MU} + S(A \vert B),
\end{equation}
where $S(X \vert B) = S(\rho^{XB})- S(\rho^{B})$ and $S(Z \vert B) = S(\rho^{ZB})- S(\rho^{B})$  are the conditional von Neumann entropies of the post measurement states

$$\rho^{XB}= \sum_{i}\left(\vert x_{i}\rangle\langle x_{i}\vert\otimes\mathbb{I}\right)\rho^{AB}\left(\vert x_{i}\rangle\langle x_{i}\vert\otimes\mathbb{I}\right),$$
$$\rho^{ZB}=\sum_{j}\left(\vert z_{j}\rangle\langle z_{j}\vert\otimes\mathbb{I}\right)\rho^{AB}\left(\vert z_{j}\rangle\langle z_{j}\vert\otimes\mathbb{I}\right),$$
  and $S(A \vert B)=S(\rho^{AB})-S(\rho^{B})$ is the conditional von Neumann entropy. We discuss some special cases: first, if measured particle $A$ and memory particle $B$ are entangled, $S(A \vert B)$ is negative and Bob's uncertainty about Alice's measurement outcomes could be reduced. Second, if $A$ and $B$ are maximally entangled then $S(A \vert B)=-\log_{2}d$ ($d$ is the dimension of measured particle). As $\log_{2}\frac{1}{c}$ cannot exceed the $\log_{2}d$, Bob can perfectly guess both $X$ and $Z$. Third, if there is no quantum memory, Eq.(3) reduces to
  \begin{equation}\
 H(X)+H(Z) \geq log_{2}\frac{1}{c} + S(A),
  \end{equation}
which is stronger than Maassen and Uffink uncertainty relation, since the measured particle is in the mixed state $S(A)\neq0$, it tightens the lower bound of Eq.(2).

  Pati \emph{et al.} \cite{Pati} proved that the uncertainties $S(X \vert B)$ and $S(Z \vert B)$ are lower bounded by an additional term compared to Eq.(3) as
\begin{align}\label{pati}
S(X \vert B)+S(Z \vert B) & \geq \log_{2}\frac{1}{c} + S(A \vert B) \\ \nonumber
& + \max\lbrace 0,D_{A}(\rho^{AB})-J_{A}(\rho^{AB}) \rbrace.
\end{align}
The classical correlation $J_{A}(\rho^{AB})$ is defined as
\begin{equation}\label{classical}
J_{A}(\rho^{AB})=S(\rho^{B})-\min_{\lbrace \Pi_{i}^{A} \rbrace}S(\rho^{B\vert \lbrace \Pi_{i}^{A} \rbrace}),
\end{equation}
where the optimization is over all POVMs ${\lbrace \Pi_{i}^{A} \rbrace}$ acting on measured particle $A$. Quantum discord is the difference between the total and the classical correlation,
\begin{equation}
D_{A}(\rho^{AB})=I(A;B)-J_{A}(\rho^{AB}),
\end{equation}
where total correlation is,
\begin{equation}
I(A;B)=S(\rho^{A})+S(\rho^{B})-S(\rho^{AB}).
\end{equation}

Lower bound in Eq.(\ref{pati}) tightens bound in Eq.(\ref{berta}) if the discord $D_{A}(\rho^{AB})$ is larger
than the classical correlation $J_{A}(\rho^{AB})$.

Coles and Piani \cite{Coles} derived an improvement on incompatibility measure, $q_{MU}$, capturing the role of the second-largest entry of $[c_{ij}]$, denoted $c_{2}$, as
\begin{equation}
S(X \vert B)+S(Z \vert B) \geq q' + S(A \vert B),
\end{equation}
where
\begin{equation}
q'= q_{MU}+ \frac{1}{2}(1- \sqrt{c})\log_{2}\frac{c}{c_{2}},
\end{equation}
when the system $A$ is a qubit then $c= c_{2}$, hence $q'=q_{MU}$.

 In this paper, we introduce a new lower bound for EUR  by adding an additional term depending on the mutual information of the bipartite state and the Holevo quantities of the ensembles that Alice prepares for Bob by her measurements.  We show that if Bob's accessible information about Alice's measurement outcomes is less than mutual information, our lower bound is tighter than the lower bound proposed by both  Berta \emph{et al.} and Pati \emph{et al}. lower bounds. We show that for complementary observables, there is a wide variety of the quantum states that for which our lower bound is stronger than the other lower bounds. We bring four examples and show that our lower bound for pure states coincides with Berta's \emph{et al}. lower bound \cite{Berta}, and for Werner states coincides with Pati's \emph{et al}. lower bound \cite{Pati}, but for Bell diagonal states and two-qubit X states our lower bound are tighter than their lower bounds. It has been found that EUR has various applications, for example in entanglement detection \cite{Partovi,Yichen,Prevedel,Li} and quantum cryptography \cite{Tomamichel,Ng}. As other applications, here we obtain a lower bound for the entanglement of formation  of bipartite quantum states and an upper bound for the regularized distillable  common randomness.

 This paper is organized as follows. In Sec.II,  we introduce  the new lower bound for EUR, and we show that for a wide variety of states, our EUR lower bound  represents an improvement to
Berta’s uncertainty relation by raising the lower bound limitis . In Sec. III, we examine our lower bound for four examples ( pure, Wernar, Bell diagonal and two-qubit X states  ), and compare the new lower bound with the other lower bounds. In Sec. IV, we discuss some of the applications of our lower bound.  Section V includes the discussion and summary of our findings.

\section{Improved Uncertainty Relation With Holevo Quantity}\label{II}

In this section, we obtain a new lower bound for EUR in the presence of memory particle. Consider a bipartite state $\rho^{AB}$ sharing between Alice and Bob.  Alice performs $X$ or $Z$ measurement and announce her choice to Bob. Bob's uncertainty about both $X$ and $Z$ measurement outcomes is
\begin{align}\label{new}
S(X \vert B)+S(Z \vert B)& = H(X)- I(X;B)+H(Z)-I(Z;B)\nonumber \\
 & \geq q_{MU} + S(A)-[I(X;B)+I(Z;B)]\nonumber \\
 & = q_{MU} + S(A|B)\nonumber \\
 & +\{I(A;B)-[I(X;B)+I(Z;B)]\},\nonumber
\end{align}
where in the second line, we apply the Eq. (4) and in the last line we use the identity $S(A)= S(A|B)+ I(A;B)$. Therefore, the EUR is obtained as
\begin{equation}\label{new1}
S(X \vert B)+S(Z \vert B)\geq q_{MU} + S(A|B)+\max\{0 , \delta\},
\end{equation}
where
\begin{equation}
 \delta=I(A;B)-[I(X;B)+I(Z;B)].
\end{equation}
 We note that when Alice measures observable P on her particle, she will obtain the i-th outcome with probability $p_{i}= tr_{AB}(\Pi^{A}_{i}\rho^{AB}\Pi^{A}_{i})$  and Bob's particle will be left in the corresponding state $\rho^{B}_{i}= \frac{tr_{A}(\Pi^{A}_{i}\rho^{AB}\Pi^{A}_{i})}{p_{i}}$,  then $$I(P;B)= S(\rho^{B})- \sum_{i}p_{i}S(\rho^{B}_{i}),$$ is the Holevo quantity and it is equal to upper bound of Bob's  accessible information about Alice's measurement outcomes. Thus, one can see that if sum of information that Alice sends to Bob by her measurements are less than the mutual information between $A$ and $B$, the above EUR represents an improvement to Berta's uncertainty relation by raising the lower bound limit by the amount of $\delta$.
 It is worth noting that the inequality Eq. (4) becomes equality if observables $X$ and $Z$ are complementary and subsystem $A$ is maximally mixed state. Thus, our lower bound is perfectly tight for the class of states with maximally mixed subsystem $A$ (including Werner states, Bell diaginal states, Isotropic states) and complementary observables. In other words, $S(X|B)+S(Z|B)$ coincides with our lower bound if $X$ and $Z$ are complementary and the subsystem $A$ is maximally mixed.

It was conjectured \cite{Sch} that the quantum mutual information is lower
bounded by the sum of the classical mutual informations in two mutually unbiased bases, namely
\begin{align}
I(A;B)\geq I(X;X')+I(Z;Z'),
\end{align}
where $X'$ and $Z'$ are two complementary observables measuring on memory particle. Although a stronger conjecture, in which $X'$ and $Z'$ are replaced by the quantum memory $B$, can be violated in general \cite{C.B}, but we will show that when $X$ and $Z$ are complementary, there are a wide variety of states for which $\delta\geq0$. We have
\begin{align}\label{evolution}
S(X \vert B)+S(Z \vert B) =& H(X)+ H(Z)- S(A) +\nonumber \\
 & S(A|B)+\delta\nonumber \\
 & \geq \log_{2} d + S(A|B),
\end{align}
where in the last line we use Berta's inequality and $d$ is the dimension of the subsystem $A$. Here we see that
\begin{align}\label{delta}
\delta \geq \log_{2} d + S(A)- H(X)- H(Z),
\end{align}
hence when the right hand side (RHS) of the above inequality is zero then $\delta\geq0$. When subsystem $A$ is maximally mixed, $S(A), H(X)$ and $H(Z)$ are equal to $ \log_{2}\ d$, making the RHS of the above equation zero. Also, when X [alternatively, Z] minimally disturbs subsystem A, H(X) [alternatively, H(Z)] is equal to S(A) and H(Z) [alternatively, H(X)] is equal to $ \log_{2}\ d$, which, again, makes the RHS zero. So, for all Bell-diagonal states, Werner states and maximally correlated mixed states we have $\delta\geq0$ and for this states our inequality tighter than Berta's \emph{et al}. uncertainty relation  Eq. (\ref{berta}). Because Pati \emph{et al}. in obtaining Eq. (\ref{pati}) put $J_{A}(\rho^{AB})$ instead of both $I(X;B)$ and $I(Z;B)$, and we know that $J_{A}(\rho^{AB})\geq I(X;B) \ \mbox{and} \ I(Z;B)$. Thus,  our lower bound is stronger than Eq. (\ref{pati}).

\section{examples}\label{III}

\subsection{Pure bipartite state}
First, we consider a pure bipartite state written in the Schmidt basis, $|\Psi\rangle_{AB}= \sum_{i}\sqrt{\lambda_{i}}|a_{i}\rangle|b_{i}\rangle$. For this state we have, $S(\rho^{A}) = S(\rho^{B})$, $I(A;B)=2S(\rho^{B})$. Alice measures observable $X$ or $Z$ on her particle. Irrelevant to which observable Alice measures, whenever she obtains a particular outcome, the state of the Bob's particle will be  pure  then $S(\rho^{B}_{i})=0$ and  $I(X;B) = I(Z;B)= S(\rho^{B})$. Thus,  $\delta=0$ and our lower bound coincides with Berta's lower bound Eq. (3).

 \subsection{Werner state}
 As a second example, we consider a two-qubit Werner state
\begin{equation}\label{spectral}
\rho^{AB} = \frac{1-p}{4}I_{A}\otimes I_{B}+p|\Psi^{-}\rangle_{AB}\langle\Psi^{-}|,
\end{equation}
where $0\leq p\leq 1$ and $|\Psi^{-}\rangle_{AB}=\frac{1}{\sqrt{2}}(|01\rangle-|10\rangle)$ is the Bell state.

 Because the Werner states are invariant under all unitary transformation of the form $U\otimes U$, so  $I(X;B) = I(Z;B)= J_{A}(\rho^{AB})$  then $\delta = \{I(A;B)-[I(X;B)+I(Z;B)]\}= D_{A}(\rho^{AB})-J_{A}(\rho^{AB})$, where we use the Eq.(7), then our lower bound equals to  which Pati \emph{et al.} introduced.

 \subsection{Bell diagonal state}
 As the third example, we consider the set of two-qubit states with the maximally mixed marginal states. This state can be written as
\begin{equation}
\rho^{AB}= \frac{1}{4}(I\otimes I +\sum^{3}_{i,j=1}w_{ij}\sigma_{i}\otimes \sigma_{j}),
\end{equation}
where $\sigma_{i}(i=1,2,3)$ are the Pauli matrices. According to the singular value decomposition theorem, the matrix $W=\{w_{ij}\}$ always can be diagonalized by a local unitary transformation, then the above state transforms to the following form:
\begin{equation}\label{State}
\rho^{AB} = \frac{1}{4}(I\otimes I + \sum_{i=1}^{3}r_{i}\sigma_{i}\otimes \sigma_{i}).
\end{equation}

 \begin{figure*}[t]\label{dd}
\begin{center}$
\begin{array}{cc}
\includegraphics[scale=.5]{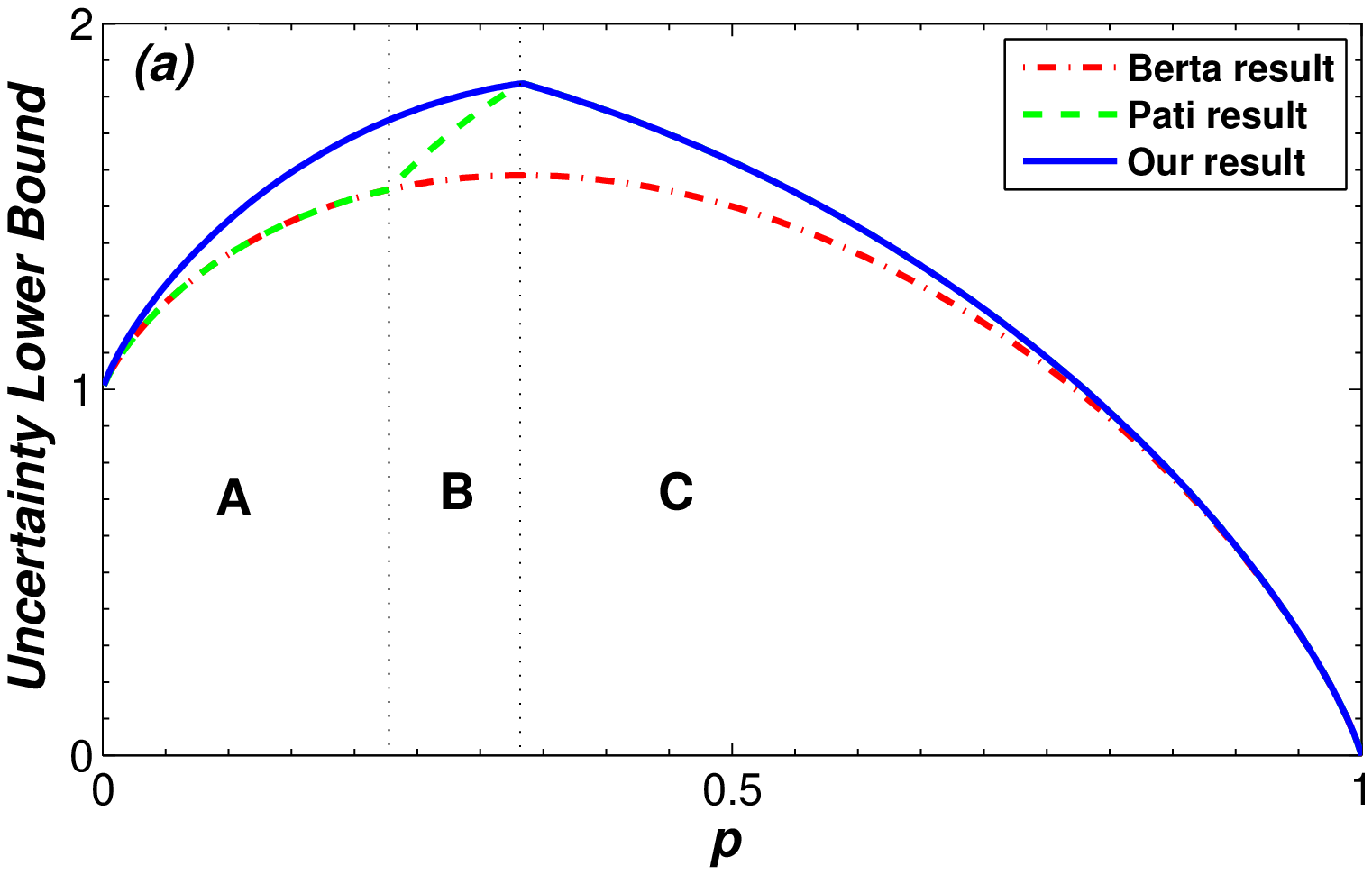}
\includegraphics[scale=.5]{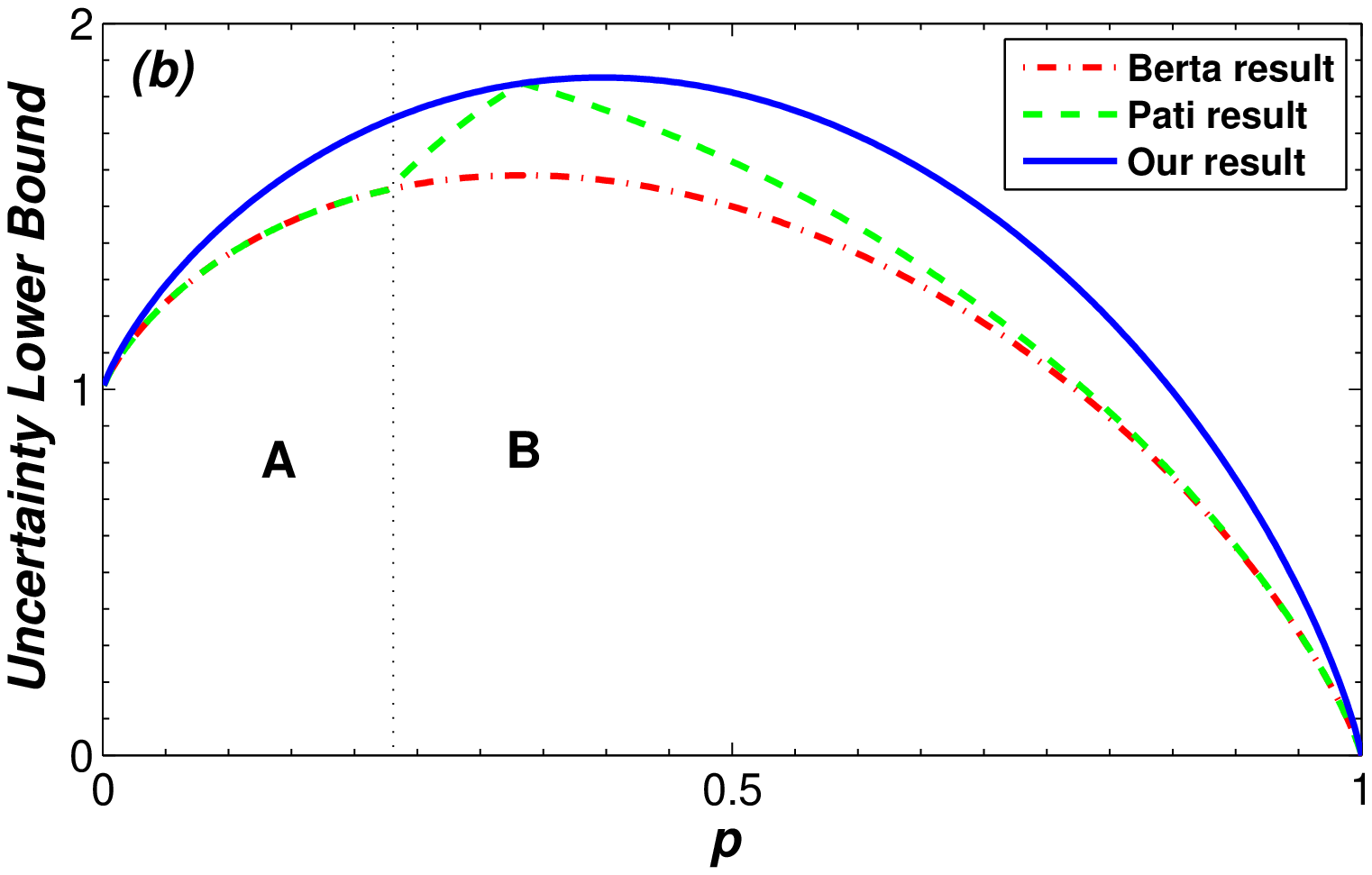}
\end{array}$
\caption{(Color online) Lower bounds of the entropic uncertainty relation of the two complementary observables in the presence of quantum memory when Bob prepare a correlated bipartite state in a special class of state: $\rho_{AB} = p|\Psi^{-}\rangle\langle\Psi^{-}|+ \frac{1-p}{2}(|\Psi^{+}\rangle\langle\Psi^{+}|+|\Phi^{+}\rangle\langle\Phi^{+}\rangle)$. The  blue (solid) line shows our results, green(dashed) line shows Pati's \emph{et al.} result and red (dot dashed Line)  represents Berta's \emph{et al.} lower bound. (a) it shows the uncertainty lower bound when one consider the two complementary observable $X$ and $Y$ i.e. choosing  $\vec{\bar{n}}=(1,0,0)$, $\vec{\bar{n}}=(0,1,0)$  respectively and (b)  shows the uncertainty lower bound when one consider the two complementary observable $X$ and $Z$ i.e.; choosing  $\vec{\bar{n}}=(1,0,0)$, $\vec{\bar{n}}=(0,0,1)$  respectively.}
\end{center}
\end{figure*}
The above density matrix is positive if $\vec{r}=(r_{1},r_{2},r_{3})$ belongs to a tetrahedron defined by the set of vertices $(-1, -1, -1), (-1, 1, 1), (1, -1, 1)$, and $(1, 1, -1)$.
A projective measurement performed by Alice can be written by $P^{A}_{\pm}= \frac{1}{2}(I\pm \vec{n}.\vec{\sigma})$ where $\vec{n}$ is a unit vector. If Alice measures observable $P$ on her particle, Bob's qubit will be in the states $\rho^{B}_{\pm}=\frac{1}{2}(I\pm \sum_{i}n_{i}r_{i}\sigma_{i})$  occurring with probability $\frac{1}{2}$. One can obtain the entropy as $$S(\rho^{B}_{\pm})=h(\frac{1+\sqrt{(n_{1}r_{1})^{2}+(n_{2}r_{2})^{2}+(n_{3}r_{3})^{2}}}{2}),$$
where $h(x)= -x\log_{2}x-(1-x)\log_{2}(1-x)$  is the binary entropy.  From $\rho^{B}=p_{+}\rho^{B}_{+}+ p_{-}\rho^{B}_{-}= \frac{1}{2}I$ and $S(\rho^{B})=1$, we conclude
$$I(P;B) = 1- h(\frac{1+\sqrt{(n_{1}r_{1})^{2}+(n_{2}r_{2})^{2}+(n_{3}r_{3})^{2}}}{2}).$$
 Now, we rearrange the three numbers $\{r_{1},r_{2},r_{3}\}$ according to their absolute values and denote the rearranged set as $\{\bar{r}_{1},\bar{r}_{2},\bar{r}_{3}\}$ such that $|\bar{r}_{1}|\geq|\bar{r}_{2}|\geq|\bar{r}_{3}|$. When $\vec{\bar{n}}=(1,0,0)$, ($\vec{\bar{n}}$ is rearranged unit vector  corresponding to $\vec{\bar{r}}$), the Holevo quantity, $I(P;B)$, reaches to its maximum, $J_{A}(\rho^{AB})=1- h(\frac{1+|\bar{r}_{1}|}{2})$. If Alice chooses $X$ such that $I(X;B)= J_{A}(\rho^{AB})$ and $Z$ is complementary to $X$, then $I(Z;B)= 1- h(\frac{1+\sqrt{(\bar{n}_{2}\bar{r}_{2})^{2}+(\bar{n}_{3}\bar{r}_{3})^{2}}}{2})$, one can see that $I(Z;B)\leq J_{A}(\rho^{AB})$, hence $\delta\geq D_{A}(\rho^{AB})-J_{A}(\rho^{AB})$ and our EUR is tighter than EUR's of Pati \emph{et al.} \cite{Pati}.

  Especially when $r_{1}=1-2p$  and $r_{2} = r_{3}= -p$, with $0\leq p \leq 1$ the state in Eq. (\ref{State}) becomes
 \begin{align}
 \rho = p|\Psi^{-}\rangle\langle\Psi^{-}|+ \frac{1-p}{2}(|\Psi^{+}\rangle\langle\Psi^{+}|+|\Phi^{+}\rangle\langle\Phi^{+}|).
 \end{align}
 Now we consider three complementary observables $X$, $Y$ and $Z$ corresponding to  $\vec{\bar{n}}=(1,0,0)$, $\vec{\bar{n}}=(0,1,0)$ and  $\vec{\bar{n}}=(0,0,1)$, respectively. One can see that

 \begin{align}
 &I(X;B)= J_{A}(\rho^{AB})=\max\{1-h(p) , 1-h(\frac{1+p}{2})\},\notag \\
 &I(Y;B)= 1-h(\frac{1+p}{2}),\notag \\
 &I(Z;B)= \min \{1-h(p), 1-h(\frac{1+p}{2})\}.
 \end{align}
The Berta's \emph{et al.}  lower bound for two sets of the complementary obsevebles $\{ X, Y\}$ and $\{ X, Z\}$ are the same and it is equal to
\begin{align}\label{BL}
q_{MU}+ S(A|B)= -p\log_{2}p-(1-p)\log_{2}(\frac{1-p}{2}),
\end{align}
and similarly the Pati's \emph{et al.} lower bound for two sets of complementary obsevebles is the same as follows
\begin{align}\label{pL}
&q_{MU}+ S(A|B)+ \max\lbrace 0,D_{A}(\rho^{AB})-J_{A}(\rho^{AB}) \rbrace=\notag \\
 &-p\log_{2}p-(1-p)\log_{2}(\frac{1-p}{2})\notag \\
 &+ \max\{(0, 2+p\log_{2}p+(1-p)\log_{2}(\frac{1-p}{2})\notag\\
 &-2\max[1-h(p), 1-h(\frac{1+p}{2})]\}.
\end{align}

The above discussion indicates that the Berta's \emph{et al.} lower bound does not able to distinguish between any two observables  in the set of the complementary observables.
In other words the lower bound is observable independent for the complementary observables. Also, this argument is true for the Pati's \emph{et al.}  lower bound.
 But, our lower bound for two sets of the complementary obsevebales  $\{ X, Y\}$ and $\{ X, Z\}$ are obtained as
\begin{align}
&q_{MU}+ S(A|B)+ \delta =  \notag \\
 &2-\max\{1-h(p), 1-h(\frac{1+p}{2}\} - h(\frac{1+p}{2}),
\end{align}
and
\begin{align}
&q_{MU}+ S(A|B)+ \delta = \nonumber\\
&2- \max\{1-h(p), 1-h(\frac{1+p}{2})\}\nonumber \\
 &- \min \{1-h(p), 1-h(\frac{1+p}{2})\},
\end{align}
respectively. As can be seen, our lower bound of EUR for two set of complementary of obsevables are different. In other words, it depends on the measured obselvables as well as correlations of quantum states. As can be seen from FIG. 1 (a) and (b), in some intervals related to parameter $p$, the results obtained by Berta, Pati and us have  overlap. In FIG. 1(a)  one consider the two complementary observable $X$ and $Y$ i.e.; corresponding to  $\vec{\bar{n}}=(1,0,0)$, $\vec{\bar{n}}=(0,1,0)$  respectively, in the $A$ region  Pati and Berta obtain the same results, However, if $p \in [1/3,1]$ then we face with the situation that our result has  overlap with Pati result  (we illustrate this by $C$ region). In FIG. 1(b) we consider  the two complementary observables $X$ and $Z$ i.e.; corresponding to  $\vec{\bar{n}}=(1,0,0)$, $\vec{\bar{n}}=(0,0,1)$  respectively. In this case, our result does not have any overlapping with the results obtained by Berta and Pati, however Berta and Pati results have overlapping in the $A$ region.

\subsection{Two-qubit X states}
 As the last example, we consider a special class of two-qubit X states

  $$\rho^{AB}= p|\Psi^{+}\rangle\langle\Psi^{+}|+(1-p)|11\rangle\langle11|$$ where $|\Psi^{+}\rangle=\frac{1}{\sqrt{2}}(|01\rangle+|10\rangle)$ is a maximally entangled state and $0\leq p\leq1$.
  The density matrices of subsystems A and B are
 $$\rho^{A}=\rho^{B}=\left(
                          \begin{array}{cc}
                            \frac{p}{2} & 0 \\
                            0 & 1-\frac{p}{2} \\
                          \end{array}
                        \right).$$

One can obtain the conditional von Neumann entropy
\begin{align}
S(A|B)=&-p\log_{2}p-(1-p)\log_{2}(1-p)+(\frac{p}{2})\log_{2}(\frac{p}{2})\notag \\
&+(1-\frac{p}{2})\log_{2}(1-\frac{p}{2}),
\end{align}
and the mutual information
\begin{align}
I(A;B)=&p\log_{2}p+(1-p)\log_{2}(1-p)-2(\frac{p}{2})\log_{2}(\frac{p}{2})\notag \\
&-2(1-\frac{p}{2})\log_{2}(1-\frac{p}{2}).
\end{align}

If Alice measures observable $X=\sigma_{x}$ or $Z=\sigma_{z}$ where $\sigma_{x}$ and $\sigma_{z}$ are Pauli matrices, then one can see that
\begin{align}
 &I(X;B)=-\frac{p}{2}\log_{2}(\frac{p}{2})-(1-\frac{p}{2})\log_{2}(1-\frac{p}{2}) \notag \\
 &+\frac{1}{2}(1-\sqrt{1-2p+2p^{2}})\log_{2}\frac{1}{2}(1-\sqrt{1-2p+2p^{2}}) \notag \\
&+\frac{1}{2}(1+\sqrt{1-2p+2p^{2}})\log_{2}\frac{1}{2}(1+\sqrt{1-2p+2p^{2}}).
 \end{align}
and
\begin{align}
 I(Z;B)=&-\frac{p}{2}\log_{2}(\frac{p}{2})-(1-\frac{p}{2})\log_{2}(1-\frac{p}{2}) \notag \\
 &+\frac{p}{2}\log_{2}(\frac{p}{2-p})+(1-p)\log_{2}(\frac{2(1-p)}{2-p}).
\end{align}
\begin{figure}[t]
\includegraphics[scale=.50]{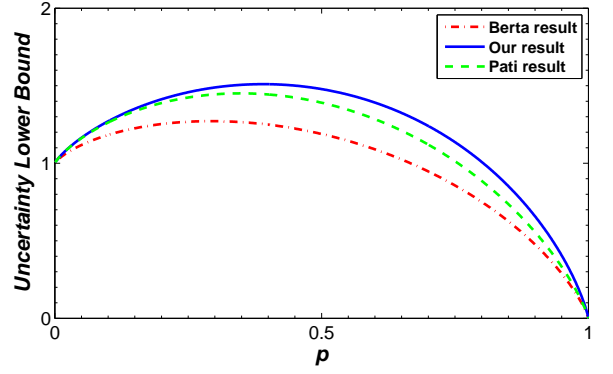}
\caption{(Color online) Lower bounds of the entropic uncertainty relation of the two complementary observables $\sigma_{x}$ and $\sigma_{z}$ in the presence of quantum memory when Bob prepare a correlated bipartite state in a special class of state: $\rho^{AB}= p|\Psi^{+}\rangle\langle\Psi^{+}|+(1-p)|11\rangle\langle11|$. The  blue (solid) line shows our results, green(dashed) line shows Pati's \emph{et al.} result and red (dot dashed ) line represents Berta's \emph{et al.} lower bound. }
\end{figure}
For this state the classical correlation equals to $I(X;B)$, therefore the quantum discord equals to $D_{A}(\rho^{AB})=I(A;B)-I(X;B)$ \cite{Fanchini,Chen,Yichen1}.
So, we can obtain the Berta's \emph{et al}. and Pati's \emph{et al.} and our lower bound. As can be seen from FIG. 2, our lower bound improves their results.

\section{applications}\label{IV}
In addition to fundamental significance, the EUR has applications in various quantum information processing task \cite{Wehner, Prevedel}. In the following we mention some of the applications. According to Eq. (\ref{berta}) if $ H(X|B) + H(Z|B)< \log_{2}\frac{1}{c}$, or  if $I(X;B) + I(Z;B)> H(X) + H(Z) - log_{2}\frac{1}{c}$, then conditional entropy $S(A|B)$ is negative and A and B must be entangled. According to our relation if $ H(X|B) + H(Z|B)< \log_{2}\frac{1}{c}+ \max\{0, \delta\}$ then the joint system is entangled. Furthermore, when $\delta\geq 0$ ;i.e. $I(A;B)\geq I(X;B) + I(Z;B)$, if $I(X;B) + I(Z;B)> S(A)$ then A and B are entangled(the conditional entropy $S(A|B)$ becomes negative), which is an improvement over using Berta's EUR.\\

Also, we can obtain a lower bound for the entanglement of formation $E_{f}(\rho_{AB})$ and  its regularized form $E_{f}^{\infty}(\rho_{AB})$. Recall that:
\begin{eqnarray}
E_{f}(\rho_{AB})=\min_{\{p_{i},|\psi_{i}\rangle\}} \sum_i p_i S(Tr_{B}[|\psi_{i}\rangle\langle\psi_{i}|]), \\ \nonumber
E_{f}^{\infty}(\rho_{AB})=\lim_{n\longrightarrow \infty} \frac{1}{n}E_{f}((\rho_{AB})^{\otimes n}), \nonumber
\end{eqnarray}
where minimum is taken over all ensembles ${\{p_{i},|\psi_{i}\rangle\}}$ satisfying $\sum_{i}p_{i}|\psi_{i}\rangle\langle\psi_{i}|= \rho_{AB}$.
 In Ref. \cite{Carlen} it was shown that $E_{f}(\rho_{AB})\geq -S(A|B)$, by using the fact that entropies are additive for tensor-power states, we conclude that $E_{f}^{\infty}(\rho_{AB})\geq -S(A|B)$. Suppose that  Alice  measures X or Z on her state and corresponding to her measurement Bob dose a measurement on his state to guess Alice's outcome. Let $P^{X}_{e}$ and $P^{Z}_{e}$ is the probabilities that Bob's guess about Alice's measurement outcomes is incorrect when she measure $X$ and $Z$ respectively. According to Fano inequality, $S(X|B) + S(Z|B)\leq b_{F}$, where $b_{F}\equiv h(P^{X}_{e}) + P^{X}_{e}\log_{2}(d-1)+ h(P^{Z}_{e}) + P^{Z}_{e}\log_{2}(d-1)$ . So, we obtain a lower bound for the  regularized entanglement of formation as follow:

\begin{equation}\label{ef}
E_{f}^{\infty}(\rho_{AB})\geq \log_{2}\frac{1}{c}+ \max\{0, \delta\} - b_{F}
\end{equation}
As an anther application, we obtain an upper bound for the regularized distillable common randomness \cite{Devetak}. Considering the n states $\rho_{CB}^{\otimes n}$ share between Charlie and Bob, then the optimum amount of classical correlation that they can share by means of classical communication form C to B, is given by
\begin{equation}
C^{\rightarrow}_{D}(\rho_{CB})=\lim_{n\longrightarrow \infty} \frac{1}{n}J(\rho_{CB})^{\otimes n}
\end{equation}
Koashi-Winter \cite{Koashi} show that

\begin{equation}
E^{\infty}_{f}(\rho_{AB})+C^{\rightarrow}_{D}(\rho_{CB})=S(\rho_B)
\end{equation}
using this equality and Eq. (\ref{ef}), we obtain an upper bound for the distillable common randomness as follow:
\begin{equation}\
C^{\rightarrow}_{D}(\rho_{CB})\leq S(\rho_B)+ b_{F} - \log_{2}\frac{1}{c}- \max\{0, \delta\}
\end{equation}

\section{Conclusion}\label{V}
We have obtained a new lower bound for the entropic uncertainty in the presence of quantum memory, by adding an additional term depending on Holevo quantity and mutual information. We have shown that  our lower bound tightens that of Berta \emph{et al}., whenever the mutual information between two particles is larger than the sum of two classical information that Alice sends to Bob by her measurements. We have demonstrated that for the complementary observables, a wide variety of the state, including Bell diagonal states and maximally correlated mixed states, fulfils this condition. We have compared our lower bound with the other lower bounds for some examples, especially for a class of Bell diagonal states and two-qubit X states, the comparison of the lower bounds are depicted in Fig. 1 and Fig. 2, which it is clear that our lower bound (blue, solid line) significantly improves the previously known results. We have discussed that the new lower bound show an improvement over the other lower bounds in entanglement detection. Using our lower bound, we have obtained a nontrivial lower bound for the entanglement of formation and an upper bound for the regularized common randomness.

\section{ACKNOWLEDGMENTS}\label{VI}
We thank the anonymous referee for helpful comments on
our paper and Mario Berta for useful discussions.








\end{document}